# Methodological monotheism across fields of science in contemporary quantitative research

Andres F. Castro Torres*, **[1]
Aliakbar Akbaritabar**

* Center for Demographic Studies, Autonomous University of Barcelona, Barcelona, Spain
([1] Corresponding author, Email: acastro@ced.uab.es)
** Max Planck Institute for Demographic Research, Rostock, Germany

## Summary

The importance of research teams' diversity for the progress of science is highlighted extensively[1–4]. Less attention has been devoted to the diversity of quantitative methods[5–8], despite the seemingly hegemonic role of hypothesis testing in modern quantitative research, epitomized by the linear model framework of analysis[9–13]. Using bibliometric data from the Web of Science[14], we conduct a large-scale and cross-disciplinary assessment of the prevalence of *linear-model-based research* from 1990 to 2022. In absolute terms, linear models are widely used across all fields of science. In relative terms, three patterns suggest linear models are hegemonic among Social Sciences. First, there is a high (>50%) and growing prevalence of *linear-model-based research*. Second, global patterns of *linear-model-based research* prevalence align with global inequalities in knowledge production[15–17]. Third, there was a citation premium to *linear-model-based research* until 2012 for articles' number of citations and for the entire period in terms of having at least one citation. Previous research suggest that the confluence of these patterns may be detrimental to the Social Sciences as it potentially marginalizes theories that are not compatible with the linear models' framework[5], lowers the diversity of narratives about social phenomena, and prevents innovative and path-breaking research[18–21], limiting the breadth of research.

## Introduction

Diversity is an asset for the practice of science. The benefits of diversity apply to research teams' composition, how phenomena are represented in concepts and theories, and how data are collected and analyzed. While research teams' diversity and size are studied extensively[1–4,22,23], the diversity of quantitative analysis frameworks has received less attention[5–7]. Given the widespread use of linear modeling techniques, a large-scale and cross-disciplinary assessment of the prevalence of linear models and their potential hegemony is lacking. In the first part of this article, we study the proportion of research articles that rely on linear models (prevalence of *linear-model-based research* herein) across six macro fields of science[24] and 254 subdisciplines [25,26]. In the second part, we study the country-level prevalence of *linear-model-based research* and compare citation patterns between articles using linear models and other statistical methods among the Social Sciences and its 47 subdisciplines. We use information from all articles indexed by the Web of Science between 1990 and May 2022[14].

Linear models are analytical frameworks where the expected value of a dependent variable is modeled as a function of independent variables, as represented by equation 1.

$$E[\mathbf{Y}] = g(\mathbf{X} \times \mathbf{b}) \qquad (1)$$

E[.] stands for expected value operator; $\mathbf{Y}$ represents a vector of length $n$ containing the values of the dependent variable; g(.) is the link function, $\mathbf{X}$ is an ($n \times p$) dimensional matrix of $p$ independent controls, mediators, or variables of interest; and $\mathbf{b}$ is a vector of length $p$ that captures the marginal associations between the independent variables and $\mathbf{Y}$. Given data and under some assumptions regarding the distribution of $\mathbf{Y}$, the distribution and correlation of the differences between the data and model's predictions, and the relationships among variables and across units of observations, the values of $\mathbf{b}$ (i.e., the regression coefficients), can be estimated and their statistical significance assessed. Different combinations of independent variables can be evaluated and compared regarding their predictive accuracy[27].

Linear models are very flexible as they can accommodate variables of different kinds (e.g., nominal, ordinal, interval, and continuous), account for grouped observations in the dependent (e.g., repeated measures models) and independent variables (e.g., hierarchical or multilevel models). Linear models can be used with multiple link functions and probability distributions[27]. Additionally, linear models' outputs are easy to summarize, present, and interpret, notwithstanding issues of misuse and misinterpretations[6,13,28–30]. This flexibility makes them very appealing for a wide variety of scientific endeavors across fields of science, from data description to causal analysis.

However, it will be hard to argue that linear models are suitable for answering all research questions within a field, for example, the Social Sciences[6,7,31,32]. Alternative data methods (*other methods* herein) have been developed, for instance, in France in the 1960s and 1970s[33,34] and the United States more recently[5,35,36]. We use the expression *other methods* to refer to all statistical techniques that do not require the specification of a dependent variable and do not aim to measure conditional associations or effects between the dependent and independent variables. Examples of *other methods* include network analysis, correspondence analysis, principal component analysis, sequence analysis, agent-based simulation, and cluster analysis. According to the promoters of these *other methods*, social realities cannot be understood by adding marginal effects. Such an approach is detrimental to a holistic understanding of social processes and relational phenomena as it ignores essential interactions, path dependencies, and structural constraints[10,37]. These critiques have been raised in various realms, including demographic change, cultural practices, and occupational trajectories, to mention a few[8,34,38].

According to previous research, the relative success of these alternative approaches in terms of the number of publications in the Social Sciences compared to linear models has been lower[5]. Research has also shown that the more significant success of linear models' application in the Social Sciences cannot be fully explained by their flexibility and suitability[6]. Historical contingencies and power dynamics among individuals and institutions have played a role in the differential success of the linear model vs. *other methods*[6,9,39,40]. As a result, several contemporary assessments of social science research suggest that the hegemony of *linear-model-based research* marginalizes specific questions and worldviews in disciplines such as economics, sociology, and demography[38,41,42].

Measuring *linear-model-based research* prevalence and assessing its potential hegemony is essential because the lack of diversity in methods (or *methodological monotheism,* as termed by Bourdieu and Wacquant[11]) could translate into a lack of diversity in narratives, standardized questions, and limited breadth of science. While the consolidation of mainstream narratives indicates increased specialization[43], it can also signal the marginalization of niche areas[44,45]. Through a gatekeeping process, *methodological monotheism* can prevent or exclude niche, disruptive, atypical, or high-risk breakthrough research endeavors deemed not conform to known traditions[7,8,18–21].

The question of whether *linear-model-based* research is hegemonic is therefore relevant for diversity, particularly for scientific fields where what constitutes evidence, explanation, and theory are socially contested and carry political ideologies and consequences, namely, the Social Sciences and the Humanities[16,17,46]. In other fields, some of these issues are less problematic. For example, a positive correlation between temperature and the length of a metal bar –measured under controlled conditions– can be interpreted in terms of a causal relationship with little to no ambiguity: the heat causes changes in the length of the bar and not the other way around[47]. In addition, observing these causes (heat) and effects (bar length) does not affect the relationship itself. In the Social Sciences, higher levels of disagreement among scholars have been associated with more significant problems' complexity[48]. Controlled conditions for observing social phenomena are rarely achieved, and when they are, results may lack external validity[12]. Therefore, mutually exclusive theories will likely co-exist with their group of supporters[32,49,50]. Social issues are often processual, path-dependent, and synergetic and may be characterized by feedback effects between causes and consequences[51]. There may also be feedback effects between describing a social phenomenon and the social phenomenon itself. These latter feedback effects arise from the inseparability of the subject and object of study[11,52].

*Linear models in the Social Sciences*

Quantitative social science research is univocally reductionist. Reducing the complexity of social realities is necessary for data collection and analysis. These complexity reductions can be implicit –i.e., inbuilt in the concept's operationalization, measurement, or data analysis– or openly discussed in the forms of methods' features, limitations, and assumptions. For example, information on individuals' educational attainment (e.g., primary, secondary, tertiary education) ignores quality differentials across institutions. At the same time, this information requires specific methods for being analyzed, namely, categorical data analysis techniques.

Linear models have grown in complexity and specialization to accommodate better specific data types (e.g., categorical variables, duration, and sequence data) and to account for data features such as nested structures, serial correlations, and complex survey sampling[53–55]. However, there are not enough discussions about the assumptions underlying the representation of the social world in the form of a linear equation[6,56]. These representations rely on assumptions regarding the nature of the relations among variables[42,56,57], the imperative to "control for," standardized, or reduce data structure to identify pure effects[58,59], and a particular understanding of causality as measurable only via quantitatively-defined counterfactuals. These counterfactuals are typically obtained using technical procedures such as randomization, matching on observable characteristics, double differentiation, and instrumental variables)[51].

These three essential features of *linear-model-based research*: (i) centered on variables, (ii) concerned with pure effects or neat associations, and (iii) conceptualizing causal relations in terms of counterfactuals, align with a concrete understanding of theory as a set of testable propositions regarding the relations between outcomes and explanatory factors. According to Abend[46], this is a valuable and legitimate definition of theory, which he terms "theory1," but it is not unique. There are at least six other uses of the word theory in Social Sciences, labeled by him as "theory2" through "theory7." Apart from Abend's "theory1," other theories are incompatible with the linear model framework. This lack of compatibility suggests that a hypothetical hegemony of *linear-model-based research* may marginalize specific theories, worldviews, and narratives about the social world, at least in quantitative Social Sciences.

To illustrate the tone, variety of fields, and the degree of discomfort with l*inear-model-based research* hegemony, we have assembled a telling set of long quotes in the supplementary material. These extracts show scholars' concerns regarding the limitations of linear model frameworks and the potentially detrimental effect of *methodological monotheism* on the Social Sciences. Explicit critiques have been put forward since the 1980s in different subdisciplines, including sociology, demography, economics, philosophy, and political sciences[6,9,38–40,53,56,60,61].

One of the most influential social scientists of the 20th century, Pierre Bourdieu, expressed it this way:

> "The particular relations between a dependent variable (such as political opinion) and so-called independent variables such as sex, age and religion, or even educational level, income and occupation tend to mask the complete system of relationships which constitutes the true principle of the specific strength and form of the effects registered in any particular correlation. The most independent of 'independent' variables conceals a whole network of statistical relations which are present, implicitly, in its relationship with any given opinion or practice"[57] [1979].

Despite being widely cited in multiple subfields in the Social Sciences, Bourdieu's idea of the centrality of 'the whole network of statistical relations' among variables and the methods he and his colleagues promoted to account for these relationships (e.g., multiple correspondence analysis), have not become hegemonic or even part of the mainstream[5,34,62]. Instead, quantitative researchers have continued, primarily searching to isolate the effects of the independent variables on outcomes, particularly in Anglophone social science tradition[63,64]. The assembled quotes in the supplementary material also demonstrate that the authors are partisan in this debate. Their concern is not only with the hegemony of *linear-model-based research* but with linear models' inferiority for examining social issues compared to the approaches and methods they promote.

We take a non-partisan approach regarding methods' suitability. Our goal is to determine empirically whether or not *linear-model-based research* is hegemonic across fields of study and subdisciplines, particularly in the Social Sciences. A method is hegemonic if three patterns coexist: (i) it is used in most articles that rely on quantitative data, (ii) its high prevalence is sustained over time, and spatial trends are consistent with existing macro-level inequalities in knowledge production, and (iii) its research has greater visibility than the research that relies on *other methods*. In light of the importance of diversity for scientific progress and innovation, we argue that all methods should be used widely, and there should not be method-related biases in citation patterns. Deviations from widespread use and equal visibility could be problematic for social science research, above and beyond methods' suitability and appropriate use, because they could reduce diversity and inspire isomorphism.

# Results

Our results come from two corpora. Corpus 1 comprises articles reporting methods or quantitative data in their abstract (7,164,784 articles). This corpus provides us with three denominators for computing the prevalence of *linear-model-based research*: (i) articles reporting any methods, (ii) articles reporting only one method, and (iii) articles reporting any methods and quantitative data. We also distinguish whether authors report linear models using general or specific terms to ensure our search terms do not drive our results. Corpus 2 includes articles using words that indicate the analysis or interpretation of empirical evidence of any kind (13,720,556 articles). Details about corpora construction are in the Data and Materials section.

***Trends in linear-model-based research prevalence across fields and subdisciplines***

Figure 1 shows time trends in the prevalence of *linear-model-based research* for the six macro fields of science. On average, more than 7,000 articles reported linear models per year in all fields except in Agricultural Sciences (3,300) and the Humanities (260). Medical and Health Sciences rank first with more than 35,000 *linear-model-based* articles per year, followed by the Natural Sciences (>15,000), the Social Sciences (>9,000), and Engineering and Technology (>7,500). In relative terms, the sustained high prevalence of *linear-model-based research* (thick red lines) suggests that these methods are hegemonic in four fields, including those dealing directly with social processes and human behavior: Agricultural Sciences, Social Sciences, and the Humanities. The two-fold significance of *linear-model-based research* for the Social Sciences in absolute and relative terms speaks to the influential role of these techniques in the field.

Figure_1

The Agricultural Sciences, Humanities, and Social Sciences display above 50% prevalence of *linear-model-based research* over the 31 years of observation among articles reporting any methods. This pattern is robust to excluding articles reporting linear models and *other methods* simultaneously, particularly in the Social Sciences, where the prevalence of *linear-model-based research* hovers around 65% for both prevalence measures. These patterns mean that under strict and more conservative estimates of *linear-model-based research* prevalence, approximately two-thirds of the research in Social Sciences has been conducted under the linear models' framework. Notably, the proportion of articles using general terms in this field (thicker green solid line) suggests that the high prevalence of *linear-model-based research* is driven chiefly by articles reporting general terms as opposed to specific models. Despite a positive trend over time, the fraction of papers reporting specific models in the Social Sciences (e.g., hierarchical models) is slightly above 10% by the end of the study period. This pattern indicates that the high prevalence of *linear-model-based research* in Social Science has been largely unaffected by the increasing specialization of methods.

The consistency of the temporal trend for the proportion of papers reporting linear models as a fraction of articles reporting quantitative data (dotted red line) and all articles in Corpus 2 (dotted blue line) suggests that the high prevalence of *linear-model-based research* among papers reporting methods in the Social Sciences is unlikely driven by the selection of articles in Corpus 1. The lower prevalence of *linear-model-based research* among articles mentioning any methods and quantitative data may be related to articles that use quantitative sources as secondary or supplementary data without performing a direct statistical analysis, which may explain the significant difference between the red solid and dotted lines in the Humanities.

In Medical and Health Sciences, more than 80% of the articles reporting methods reported linear models. We interpret this pattern as unsurprising and indicative of the appropriateness of our measurement strategy because medical and health subdisciplines often deal with randomized control trials and clinical trials data, or research questions that involve identifying causal relationships between medical treatments and individuals' health or policy interventions and populations' well-being[65]. Linear models are well suited for these types of standardized questions.

The patterns observed for Agricultural Sciences, Humanities, Medical and Health Sciences, and Social Sciences contrast with the low prevalence of *linear-model-based research* in the Engineering and Technology, and Natural Sciences. Notably, for our measurement strategy, the proportion of papers reporting any methods is higher or equal in these two latter fields (0.28 and 0.22) compared to the Social Sciences (0.22), which suggests that field-specific trends in the prevalence of *linear-model-based research* are unlikely driven by our selection of terms. Likewise, the lowest proportion of papers reporting methods in the Humanities (0.09) is consistent with the lower affinity of these subdisciplines and statistical analyses. In addition, observed trends in Engineering and Technology, and Natural Sciences could have been driven by what is considered "methodological pluralism" in literature[8] since these fields have an inductive approach to science and would use a more diverse multitude of observational and analytical methods.

Figure 2 shows the scatter plot between the prevalence of linear models as a fraction of papers reporting any methods and the annual average percentage point change in the prevalence of linear models from 1990 to 2022 for most subdisciplines across macro fields of science. The right panel zooms into the areas of the plot where most of the Social Sciences subdisciplines cluster. The overarching pattern in this figure confirms that patterns documented in Figure 1 are valid for most of the subdisciplines in the Social Sciences, particularly those with a significant number of indexed articles (number of articles > 50,000).

Figure_2

According to the left panel in Figure 2, the relative position of the *Statistics & Probability* (49% articles reporting linear models and -0.54 p.p. Annual average change in the proportion of papers reporting linear models) further our confidence in our measurement approach. This subdiscipline comprises the articles that develop most of the applications of statistical methods to other subdisciplines; its middle-point position on the proportion of papers reporting linear models indicates our corpora and measures are appropriately selected and defined. In addition, this suggests that statisticians work and develop linear models and *other methods* evenly. This panel also shows that only five of the 47 subdisciplines in the Social Sciences display linear model prevalences below 50%: *Asian Studies*; *Cultural Studies*; *Information Science & Library Science*; *Operation Research & Management Sciences*; and *Psychology, Mathematics*.

As shown in the right panel of Figure 2, 42 of the 47 Social Science subdisciplines, including several with more than 50,000 indexed articles (labeled), display a prevalence of *linear-model-based research* above 50%; 34 of them display positive temporal trends meaning that *linear-model-based research* grew over the past 31 years. For example, *Education & Educational Research* displays a linear model prevalence of 70% along with a 0.72 average percentage point yearly increase from 1990 to 2022, i.e., more than 22 percentage point absolute increase in the prevalence of *linear-model-based research* during the period of analysis. The clustering in high and growing prevalences of *linear-model-based research* among Social Sciences subdisciplines is only comparable to that of the Medical and Health Science, although the latter display a much higher prevalence of *linear-model-based research* and relatively lower yearly average increase due to ceiling effects.

***The North and South patterns of linear-model-based research hegemony in the Social Sciences***

As seen in the top panel in Figure 3, the countries of the global North (i.e., Northern America, West and Northern Europe, and Australia) and China contribute the largest share of papers in the quantitative Social Sciences indexed in WOS (Corpus 2). The US is the most significant contributor with exp(-0.96) = 38% of the total papers in our sample, followed by the UK (7.7%), China (5.7%), Germany (4.1%), and Canada (4.1%). The five remaining countries of the top ten producers -Australia, Spain, the Netherlands, Italy, and France, contribute between 1.9% and 3.9%. Together, these top ten producers account for *almost two-thirds* of all articles in our sample (74.7%). These global patterns suggest that our corpora, dominated by English-speaking literature from Western, educated, industrialized, rich, and democratic (WEIRD) countries[66–68], are not biased regarding the geographical origin of articles[15,69].

Figure_3

The country-level pattern of *linear-model-based research* prevalence (bottom panel in Figure 3) is partially consistent with the global inequalities and power relations in knowledge production documented in the top panel and elsewhere[15]. Center-periphery ties, where the center exerts dominance in setting up research agendas and research methods, are typically observed in the global dynamics of knowledge production[16,70–73]. For example, countries with the lowest share of papers in Sub-Saharan Africa (periphery) display the highest *linear-model-based research* prevalence. These countries are followed by the most prominent producers (i.e., the global North), which, given their share of articles, drive the overarching trend of *linear-model–based research* hegemony (center). The US is the most prominent case, with 38% of the total papers and 74% reporting linear models among papers reporting any methods.

France displays a distinctive pattern among the top ten producers and Western Europe: relatively large shares of Social Sciences articles (1.8%) and a relatively low prevalence of *linear-model-based research* (49%). This geographic exception may be related to the development of the French school of statistics and its relation with the Social Sciences produced in this country's research institutions. French-speaking quantitative Social Sciences were highly influenced by Pierre Bourdieu, who worked closely with statisticians in applying the today termed Geometric Data Analysis techniques (GDA) to the Social Sciences[34,62,74]. To a large extent, these techniques were conceived in opposition to regression techniques, which may have contributed to a lower prevalence of *linear-model-based research* relative to other European countries and the US[33,62,75].

The French exceptionalism, the overarching North and South differences in the prevalence of *linear-model-based research*, and the concentration of articles in a handful of English-speaking countries (the US, Australia, and the UK), all from the global North, signal a broader opposition between Anglo and non-Anglo tradition of quantitative analysis in the Social Sciences[75]. For example, Eastern countries with a large share of the research papers, such as Iran, India, and China, display a low prevalence of *linear-model-based research*. The fact that these three countries are from the East suggests that their lower prevalence of *linear-model-based research* can be interpreted as a more moderate influence of Anglophone traditions of quantitative data analysis compared to other relatively big producers in the West, such as Brazil and Chile in Latin America, or several Eastern European countries where US institutions may have exerted a more substantial influence. While trends in France and Eastern non-English speaking countries could be considered as driven by the fact that their scientific publications could be in non-English languages, the observed trends in Latin America and African non-English speaking countries and the significant share in the production of Social Sciences articles (top panel in Figure 3) indicates that our corpus is not biased by geography or language.

*The greater visibility of linear-model-based research versus other types of research*

Citation patterns support the hypotheses of higher visibility of *linear-model-based research* (i.e., a citation premium) compared to articles reporting *other methods* and quantitative data. From 1990 to the present *linear-model-based research* was consistently more likely to have at least one citation than other quantitative papers, particularly among countries in the top ten producers. In addition, from 1990 to 2012, *linear-model-based research* received, on average, more citations than studies reporting *other methods*. The last six years of observation (2013-2018) display a reversal of the *linear-model-based research* premium in the average number of citations.

Figure 4 displays the association between reporting linear models and the odds of receiving at least one citation (left panel) and the average number of citations (right panel) within the three years after publication. Positive values indicate a *linear-model-based research* citation premium compared to research reporting *other methods* (red dots and lines) and research reporting *other methods* or quantitative data (light green dots and lines). Results are stratified by the top ten producer countries, world regions, and for two periods: 1990-2012 and 2013-2018. And we control for whether the article is from a single country (vs. multiple countries) and the single year of publication.

Figure_4

In line with the third aspect of our definition of hegemony (higher visibility) and the spatial patterns documented in the previous section, citation premiums are at play in the centers of academic production (i.e., countries of the global North). According to the left panel in Figure 4, from 1990 to 2012 (empty circles), there was a positive association between reporting linear models and receiving at least one citation, particularly for articles from the top ten producer countries. Exceptions to this pattern include some top countries where the association is positive, but the confidence interval contains zero (the UK, Spain, and Italy), and some regions where the association was negative (Central and Southern Asia, North Africa and Western Asia). Notably, the only positive and significant coefficient across regional groups pertains to Europe and Northern America, which speak to the differential value of methods between the center and peripheral countries. From 2013 to 2018 (filled circles), this visibility premium was held in half of the top ten producer countries: the US, China, Australia, the Netherlands, and France, and weakened elsewhere; only Germany and Italy displayed negative coefficients for 2013-1018, and their confidence intervals include zero suggesting there was neither premium nor penalty. Across regions, virtually all coefficients became negative (except for Sub-Saharan Africa), further highlighting the differential value of methods between top producers (center) and the rest of the world (periphery).

The average number of citations mirrors these patterns. According to the right panel in Figure 4, during the first two decades of analysis (1990-2012), articles reporting linear models received, on average, more citations than articles reporting *other methods* (red dots and lines). This is true for all top ten producers except China and Spain. For example, on average, articles from the US reporting linear models had 10.7% more citations (coefficient = 0.1) than articles reporting *other methods*. This relative advantage is relevant for the absolute number of citations, given the magnitude of the geographical disparities in the share of articles (top panel in Figure 3) and the high prevalence of *linear-model-based research* among articles reporting any methods (bottom panel in Figure 3). These patterns of greater visibility for *linear-model-based research* before 2012 are consistent when the reference group comprises articles reporting *other methods* or quantitative data (light green markers). This consistency means that even after accounting for potential methods reporting biases, *linear-model-based research* was associated with greater visibility than other quantitative research, more broadly defined.

From 2013 to 2018, the above-mentioned *linear-model-based research's* citation premium reversed among the top ten producer countries and became more negative among regional groups. All the filled red dots lay on the negative side of the plot, and only the confidence intervals for Sub-Saharan Africa include zero. These negative and statistically significant associations mean that during the last seven years of observation, *linear-model-based research* was associated with a citation penalty, which signals a potential reversal in their hegemony, at least regarding research visibility in both the center and the periphery.

To understand the potential drivers of this penalty, we replicated the right panel of Figure 4, excluding the top 10% and 5% articles in the number of citations within each country and region (see Figure A2). These two replications show no citation penalty for any of the top ten producer countries. Indeed, the *linear-model-based research* citation premium is observed for the US, China, and Australia when the top 5% and 10% of cited articles are excluded from the sample. The citation penalty was held for all regions, although confidence intervals include zero for Europe, Northern America, and Sub-Saharan Africa. These patterns indicate that the reversal in the *linear-model-based research* premium may be driven by a few articles (produced in top ten producer countries) using *other methods* and receiving a disproportionately large number of citations.

## Conclusion

We define and implement measures of *linear-model-based research* prevalence using large-scale datasets on all articles (40+ million) from the Web of Science (WOS) covering 1990-2022 and six macro fields of science. The first part of our analysis validates our measurement approach and documents substantial cross-field heterogeneity in the prevalence of *linear-model-based research*. In absolute terms, the average number of articles reporting linear models ranged from a few hundred in the Humanities to more than 35,000 in Medical and Health Science. The Social Sciences ranked third, with almost 9,000 articles per year. In relative terms, the prevalence of linear-model-based research ranged from low levels in Engineering and Technology (20%) to very high levels in Medical and Health Sciences (80%). The Social Sciences ranked second, with *linear-model-based research* prevalences hovering around 65% for the entire analysis period.

The second part of our analysis confirms that in the Social Sciences, *linear-model-based research* is hegemonic. We conclude this based on three confluent patterns: (i) the high, sustained, and growing prevalence of linear models over time in Social Sciences in general and across its subdisciplines; (ii) geographical patterns in *linear-model-based research* prevalence that are consistent with global inequalities in knowledge production; and (iii) the existence of a citation premium that has favored *linear-model-based research* at least until 2012 for the number of citations, and for the entire period in terms of having at least one citation (i.e., avoiding invisibility). Despite being incapable of establishing causal relationships among the processes underlying these patterns, the salience of our results allows us to speculate regarding the potential drivers and implications of these patterns. We rely on some previous, in-depth, single-discipline analyses[5–7] and historical accounts of scholarly traditions' developments[9,39,62,75] to inform our speculations and suggest future research areas.

Several scholars have cautioned us regarding the limitations of linear model approaches and the need to consider their presumptions[6,38,42,56,76]. These limitations would not be a concern if linear models were used as much as *other methods*. However, some discipline-specific studies have shown how linear modeling and hypothesis testing became normative in journals and fields such as economics, demography, sociology, psychology, and political sciences[6,9,13,38,40,64,77]. Our study extends these results to four macro fields of science and more in-depth to 47 subdisciplines in contemporary quantitative Social Sciences by documenting a sustained, high, and growing prevalence of *linear-models-based research* over the past 31 years. We termed this confluence of patterns *methodological monotheism*. This *methodological monotheism* and social scientists' discomfort with linear model frameworks suggest that we may be neglecting or entirely missing perspectives and approaches that do not conform with this analytical framework and therefore losing opportunities to expand our understanding of the social world and its problematics.

Studies on the use of quantitative methods in Sociology have examined the institutional- and individual-level determinants of authors' use of alternative quantitative methods[5]. According to these studies, conforming with mainstream analysis methods may be more beneficial for individual careers because existing institutional mechanisms reward this type of conformism and punish (perhaps non–intentionally) *other methods*. Only authors with academic authority, seniority, prestige, and institutional credentials can afford the risk of exploring a non-mainstream perspective. And yet, their success in spreading the use of a given method is not always warranted, as shown by the case of correspondence analysis and qualitative comparative analysis[5]. We cannot look at the institutional- and individual-level determinants in all the 47 subdisciplines we study; however, the similarity among patterns between our results and previous in-depth single-discipline studies suggests that analogous mechanisms may be at play across these social science subdisciplines.

The spatial patterns of linear models' prevalence indicate that *methodological monotheism* results from the combination of high yet distinct prevalences: very high (>90%) in peripheral countries, high (80% to 90%) in the chief countries of Social Science production (center), intermediate (50% to 79%) in semi-periphery countries, and relatively low among regional leaders in non-Anglo-speaking areas of the East (e.g., Turkey, China, India, and Iran). These patterns are consistent with the idea that knowledge production can be described as center-periphery relations between the global North and South, the East and the West, and Anglo- vs. non-Anglo-speaking social science traditions[71–73].

The documented patterns suggest that the centers of knowledge production in the global North (primarily Anglo-speaking countries) exert a double influence by tacitly imposing the mainstream framework for quantitative analysis (the driving force of aggregated *methodological monotheism*). At the same time, specific institutions and researchers in the West/North[78] can afford to risk applying and promoting alternative frameworks, despite not being entirely successful[5]. The further distance to the center, the higher the value risk of departing from the mainstream; therefore, the greater prevalence of *linear-model-based research*. Some specific regional exceptions, such as China, Turkey, Algeria, and Iran, may be explained by the differential development of the scholarly tradition in the Westernization of academia. For example, divergent patterns in *linear-model-based research* hegemony between countries like Brazil, on the one hand, and China, India, and Iran, on the other, may be explained by the differential influence of Anglo-speaking traditions of analysis and the differential need to conform to mainstream approaches. Historical non-Anglo traditions may have created sufficiently large academic communities such that hegemony, as defined here, is more difficult to establish. These patterns are despite increasing and significant scientific collaborations between global North countries and the rest of the world, meaning that, although partnerships are essential, they are not associated with a unique pattern of *linear-model-based research* prevalence.

Linear models hegemony in terms of use, country-level distribution, and citation patterns is consequential in several ways. **First**, hegemony could preclude the emergence, widespread use, and extension of potential path-breaking, atypical or disruptive approaches that could better solve current societal problems (e.g., rising social inequality, climate change, increased vulnerability of minorities). **Second**, global inequalities in knowledge production may be reinforced by *methodological monotheism* as the capacity to develop new methods and bring them to the forefront of research is not evenly distributed across countries and institutions. Just as individuals with more extraordinary credentials and prestige can afford the risk of path-breaking analysis, countries, and regions in privileged positions[79] (e.g., highly funded institutions) can lead research under alternative (risky) approaches[1,8,80–82] without strategic concerns for publication or research evaluation[83,84]. Suppose the linear model's greater visibility is to disappear due to successful innovations in data analysis techniques outside the linear models' framework (as suggested by the recent turn of the *linear-model-based research* premium and despite the sustained invisibility of the *other methods*). In that case, country- and institutional-level inequalities in visibility are likely to increase, favoring the innovation leaders[85,86]. **Third**, to the extent that *methodological monotheism* also exists in research training programs and institutions, it may take some decades to disappear because, although the visibility premium has reversed, many generations of researchers and instructors were trained under a context that privileges *linear model-based research*.

Our study carries several limitations beyond its very descriptive nature. We cannot evaluate other aspects of hegemony, including potential bias in the methods taught in graduate training programs, editors' and reviewers' preferences for certain types of analysis over others, or the concrete reasons for which researchers across different contexts decide in favor or against a given method and influence others decisions[87]. In addition, our data cover only recent decades, when hypothesis-testing and falsification practices may be positively rewarded as part of the practice of normal science, to use Kuhn's terminology[13,88].

There are more comprehensive text analysis methods to identify all noun-phrase-clauses or word combinations that our list might have overlooked (e.g., by selecting an anchor term and finding adjectives pairing up with it). Nevertheless, we decided on the most straightforward and strict method of search, aiming to find the lower bound of hegemonic use of techniques. Using more complex text analysis methods will likely increase the observed trends in the prevalence of *linear-model-based research*.

## Data and Materials

We use publications metadata from 40,603,923 articles indexed by Clarivate's Web of Science (WOS) provided by the German Competence Centre for Bibliometrics[89] via Max Planck Digital Library. We excluded all research articles with no abstract. WOS is one of the most exhaustive bibliometric databases covering more than 21 thousand journals[14,90]. We limit our analysis to *Article* document type. We focus on abstracts because the terms we investigate here are more likely to be reported in this part than in titles or keywords. We know abstracts could not include all the details[91–93]. We know that WOS is over-represented by English language journals and WEIRD countries[66–68]. This bias implies that despite the presented trends of social science production being dispersed in diverse geographical areas and non-English speaking countries, nevertheless, our results concern a specific area of the existing research, mainly produced under the global North country standards. We select two corpora from the WOS data.

*Corpus 1* comprises all research articles where abstracts include at least one term referring to quantitative data (e.g., survey, census, administrative records) or statistical methods. The terms referring to statistical methods are organized according to whether they refer to a linear model (e.g., regression analysis) or *other methods* (e.g., correspondence analysis). In addition, each term is classified as general or specific depending on whether they use a generic or specific name for describing the methods (e.g., regression vs. logistic regression). We built an initial list of 57 terms. We circulated this list among six colleagues with diverse disciplinary backgrounds, including health sciences, natural and social sciences, and statistics, asking them to add missing methods or comment on the ones already included. After a few exchanges and clarification with colleagues, we consolidated a list of 73 terms; this list was further extended to 163 while conducting the literature review and adding both hyphenated/non-hyphenated names and American and British spelling conventions. Of these 163 terms, 15 refer to quantitative data, 110 refer to linear models (48 general, 50 specifics, 12 causality-related words), and 38 refer to *other methods* (18 general, 20 specific). The complete list of terms and their classification are available in the Supplementary Material. This corpus includes 7,164,784 articles.

*Corpus 2* comprises research articles where abstracts include *one of* the following seven words: 'model,' 'data,' 'evidence,' 'empirical,' 'results,' 'method' or 'methods.' **and** *one of* the following five words: 'analysis,' 'analyze,' 'analyse,' 'study,' or 'investigate' used once or more in the same abstract. We argue that these keywords allow us to identify *potentially empirical papers* dealing with processing, analyzing, or interpreting data. Yet, qualitative and other studies can be included in this corpus, although they are not part of the *ideal risk set*. This corpus includes 13,720,556 articles.

*Prevalence measures:* We compute four different measures of *linear-model-based research* prevalence. Our primary measure of interest is the proportion of articles that report linear models in their abstract among all articles reporting any statistical method. In a second prevalence measure, articles that report linear models and *other methods* simultaneously are subtracted from the first *linear-model-based research* prevalence. Third, we compute the prevalence of linear models using the total number of articles reporting methods or quantitative data sources as the denominator. Finally, we compute the prevalence of *linear-model-based research* among all articles in *Corpus 2*. These latter two measures of prevalence help assess the consistency of our results over time from a more conservative and strict perspective, i.e., by enlarging the ideal risk set. To evaluate whether our results are driven by reporting general or specific terms, we report the share of articles using these two types of words among papers reporting linear models (See supplementary Information for figures separating the use of these term groups).

*(In)visibility measures:* We compute two measures of papers' visibility: the total number of citations received in the three years following publication and whether the paper received at least one citation during the same period. The three-year window time warrants comparability between older and more recent papers, and it is justified because articles' citations mature in the third year after publication considering disciplinary differences[94,95]. At the same time, it could penalize articles published in later months of the same year[96].

*Measuring linear-model-based research relative visibility:* To examine *linear-model-based research's* advantage in visibility outcomes, we fitted a series of negative binomial (link function = log) and binomial (link function = logit) models predicting our two visibility measures. The negative binomial distribution is suitable for modeling the number of citations in the first three years, given the strongly skewed distribution of this outcome. The binomial distribution is appropriate for modeling whether articles receive at least one citation. Our primary variable of interest in these models is whether the authors reported or not a linear model in their abstract. Hence, the regression coefficients of our variable of interest measure the difference in the average number of citations and the odds of having at least one citation between *linear-model-based research* and quantitative research that report *other methods* or quantitative data.

To account for geographical differences in citation patterns, we stratified our models by the top ten countries regarding the number of articles and the rest of the countries grouped into the United Nations Sustainable development regions[97]. To assess potential changes over time in *linear-model-based research* visibility advantage, we ran separate models for articles published before and after the median year of publication (i.e., 2012). This partition favors the symmetry in the sample size and, therefore, the uncertainty of estimates in each sample. We control for the year of publication, and whether an article involves or not authors from different countries, i.e., it is a product of international collaboration. These control variables are essential because the number of citations and the proportion of papers without citations increases and decrease over time, respectively[86,98,99]. And articles involving authors from only one country receive, on average, fewer citations and are more likely to have zero citations[85,100,101].

To categorize publications into fields of science[24], we use a mapping of OECD macro fields of science to WOS subject classifications. Because the field of science categories are not mutually exclusive, articles can be included in more than one category at a time. We allow this multiple counting for field-specific analysis (Figures 1 and 2). Instead, we consider unique articles (single counting) in our country-level analysis and statistical models, which only deal with social science publications. In addition, we include only different countries per article (e.g., if an article has multiple authors from the same country and one author from another country, we consider it a product of the two countries and use distinct country addresses per article to account for international versus single country publications).

**Author contributions**
AFCT: Conceptualization, Methodology, Software, Validation, Formal analysis, Investigation, Writing - Original Draft, Writing - Review & Editing, Visualization
AA: Conceptualization, Software, Validation, Data Curation, Investigation, Writing - Original Draft, Writing - Review & Editing

**Acknowledgments**
We thank our colleagues Diego Alburez-Gutierrez, Enrique Acosta, Maarten Jacob Bijlsma, Beatriz Sofia Gil, Esther Dorothea Denecke, and Robert Gordon Rinderknecht, who reviewed and complemented the list of terms we have used here. We would also like to thank Diego Alburez-Gutierrez and Misha Teplitskiy for their helpful comments on the draft of our manuscript.

**Funding Acknowledgement**
This study has received access to the bibliometric data through the project "Kompetenzzentrum Bibliometrie" and we acknowledge their funder Bundesministerium für Bildung und Forschung (grant number 01PQ17001).

**Data Availability**
We use data from the German Competence Centre for Bibliometrics (grant number 01PQ17001). Restrictions apply to the availability of these data, which were used under license for the current study and are not publicly available.

**Fig. 1**

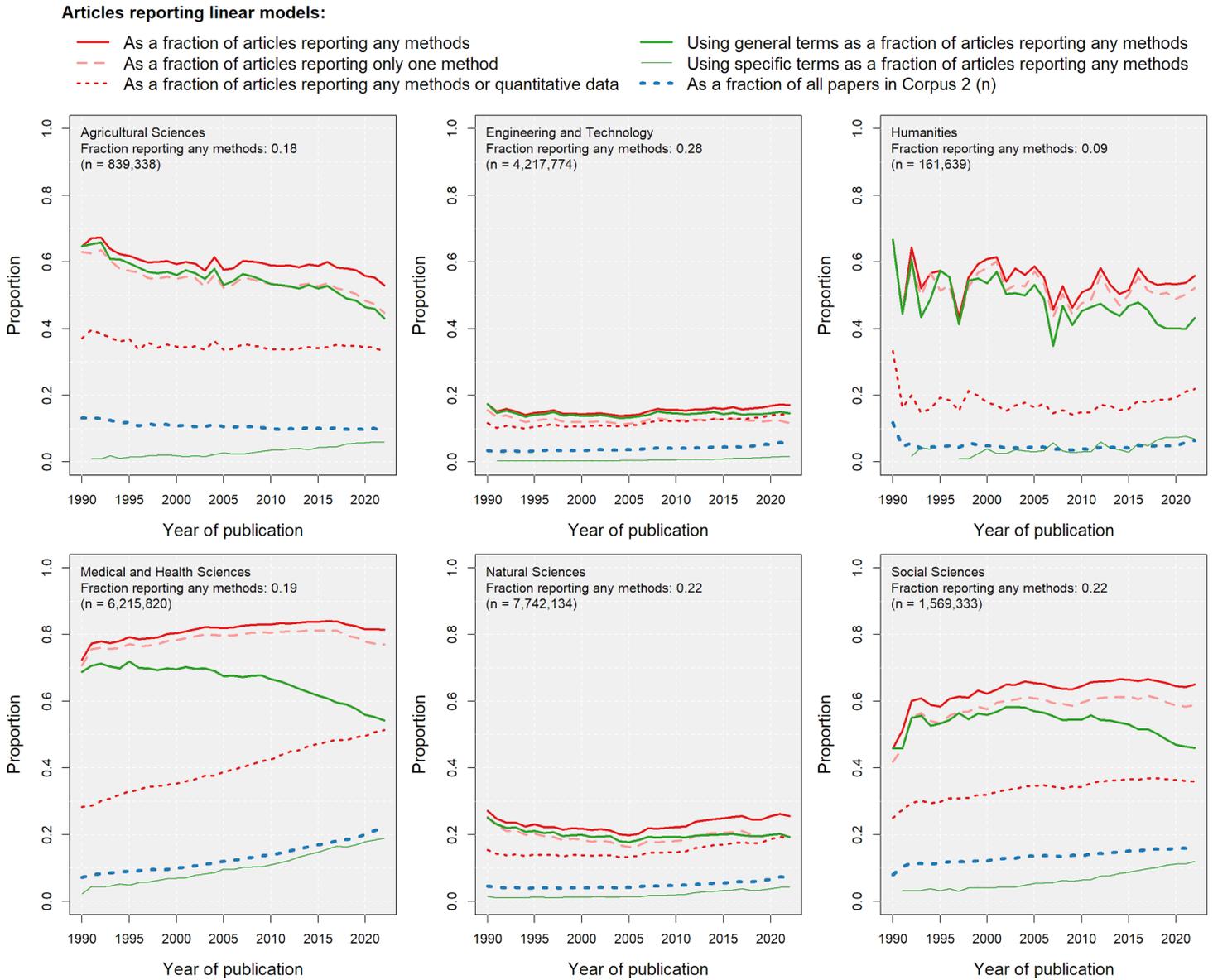

**Sustained high prevalence of *linear-model-based research* in four of six macro fields of science.** Time trends in four measures of the prevalence of *linear-model-based research* (red lines) and relative distribution of general (thicker green solid line) and specific (thinner green solid line) terms for reporting linear models in articles' abstracts across the six OECD fields of science, 1990 to 2022.

**Fig. 2**

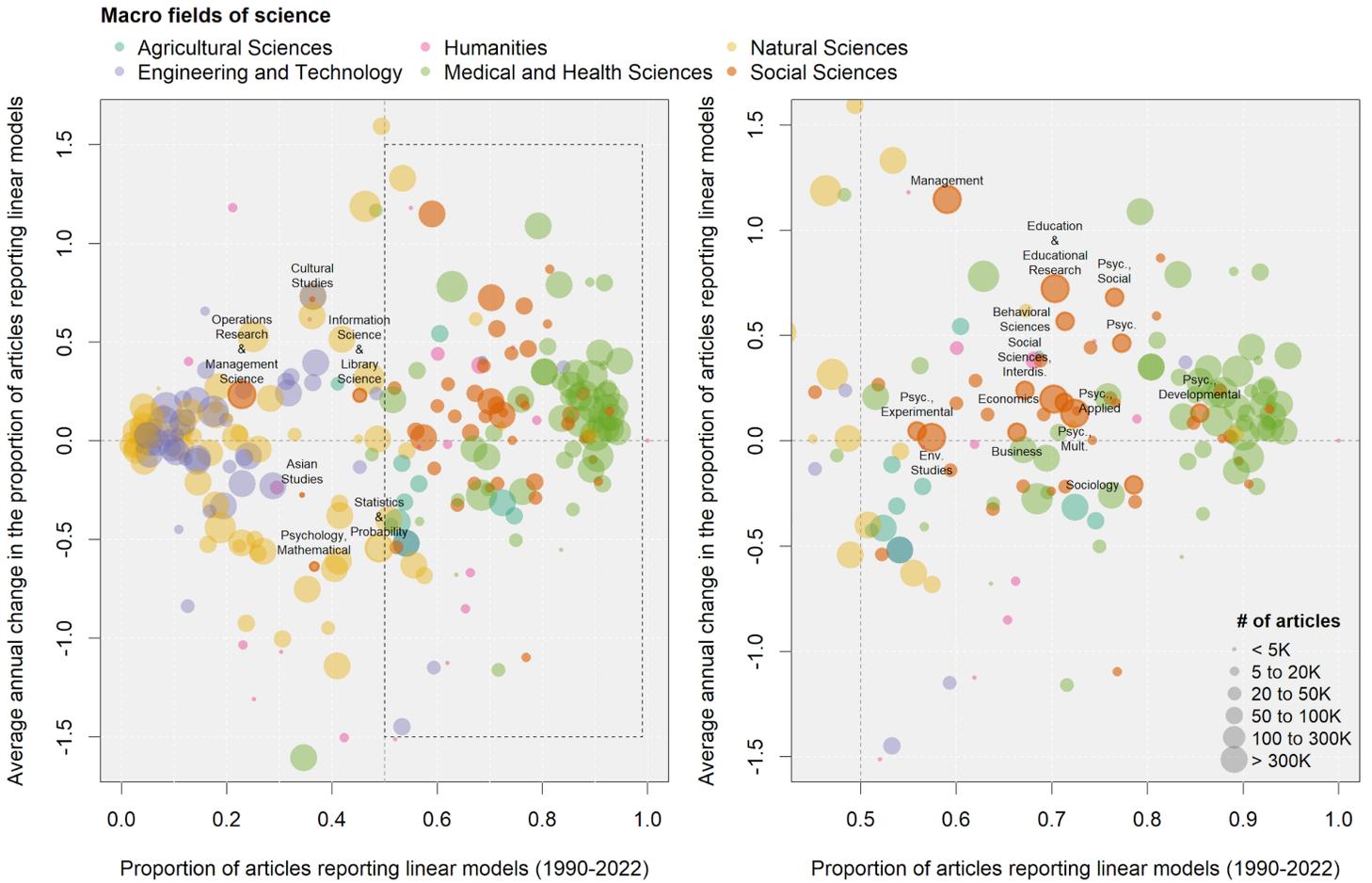

**Social Science subdisciplines cluster above 50%, and growing prevalence of *linear-model–based research*.** Distribution of subdisciplines according to the prevalence of *linear-model-based research* and the average yearly change in the prevalence of *linear-model-based research* as a fraction of articles reporting any methods from 1990 to 2022. *Statistics & Probability* and Social Sciences subdisciplines with more than 50,000 articles are labeled.

**Fig 3.**

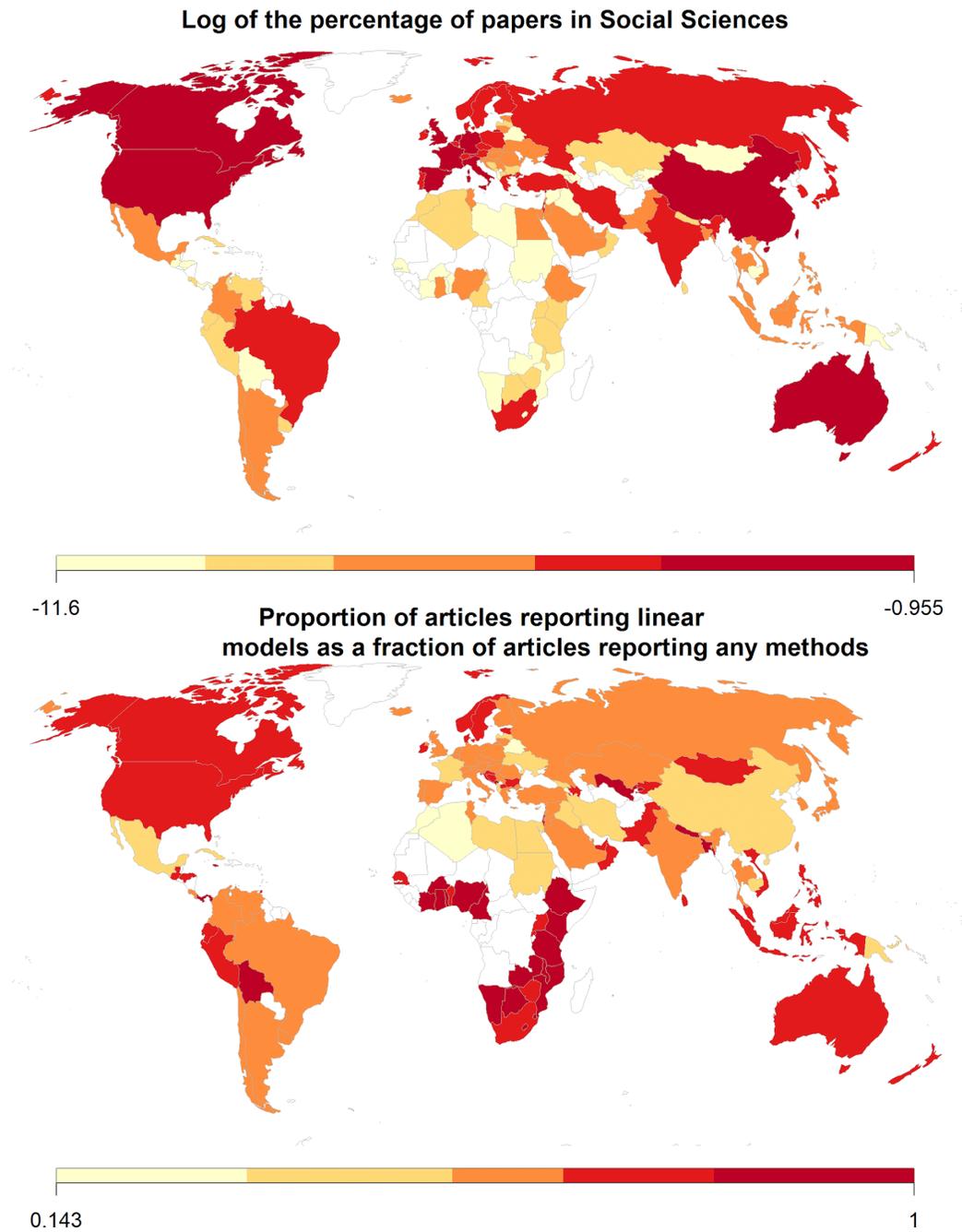

**Global patterns of Social Science production and linear models hegemony, 1990-2021.** Country-level share of articles in Social Sciences (top panel) and prevalence of *linear-model-based research* for single-country articles (bottom panel). Information for countries with less than ten articles in the WOS is ignored. Cut-off points for the top panel are the 25th (0.006%), 50th (0.029%), 75th (0.35%), 90th (1.68%), and 100th (38.5%) percentiles. Cut-off points for the bottom panel (obtained via Jenks' algorithm) are 0.14, 0.33, 0.54, 0.65, 0.80, and 1.00. This selection of cut-off points assigned the top ten countries on each scale to the first category and favors the readability of the remaining ones.

**Fig 4.**

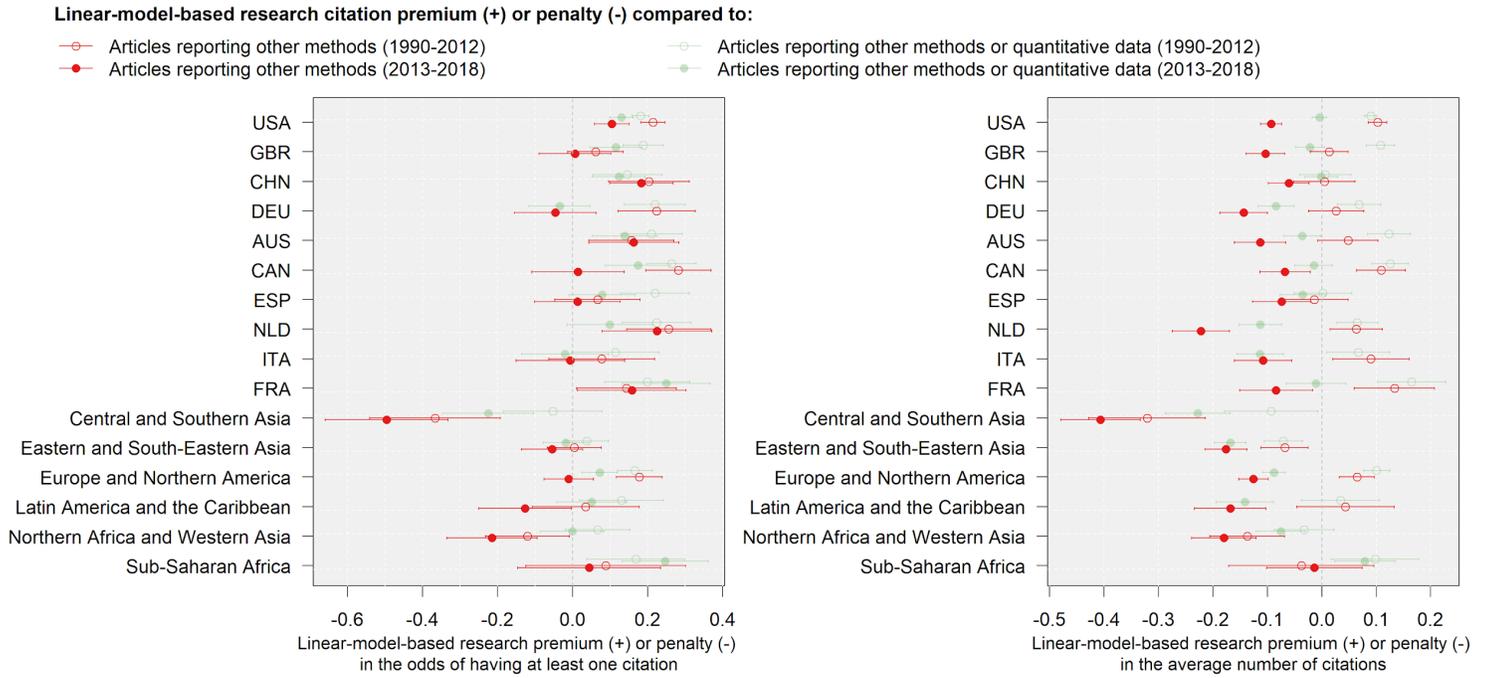

**The visibility advantage of *linear-model-based research* compared to quantitative research using *other methods* and quantitative data sources.** *Linear-model-based research* premium and penalty in the odds of having at least one citation (left panel) and the average number of citations (right panel) in the first three years after publication between (i) articles reporting linear models and articles reporting *other methods* (red dots and lines) and (ii) articles reporting *other methods* or quantitative data (light green dots and lines). Regressions are stratified by countries with the ten largest shares of articles in the Social Sciences (USA: the United States of America, CHN: China, GBR: the United Kingdom, DEU: Germany, NDL: the Netherlands, FRA: France, CAN: Canada, ITA: Italy, ESP: Spain, and AUS: Australia) and the UN-SDG regions for the other countries and by the median year of publication (i.e., 2012). Australia and New Zealand are included in Europe and Northern America, and other countries from Oceania are excluded due to the small sample size (1,350 articles).

**Supplementary material**

**This file includes**:
- A. Quotes expressing scholars' discomfort with the methodological presumptions of regression analysis and linear-model-based research in different domains
- B. List of terms for corpora construction
- C. Finding linear model-based terms in publication's abstracts
- D. Complementary results and figures

*A. Quotes expressing scholars' discomfort with the methodological presumptions of regression analysis and linear-model-based research in different domains*

One of the earliest and most widely cited critiques of *linear-model-based research* was expressed by Andrew Abbott in the late 1980s. Our selected quote from his text expresses the 'blinding' effect that the general linear reality assumption is causing to the social sciences.

> "I shall argue that there is implicit in standard methods a "general linear reality" (GLR), a set of deep assumptions about how and why social events occur, and that these assumptions prevent the analysis of many problems interesting to theorists and empiricists alike. In addition to delineating these assumptions, I shall consider alternative methods relaxing them. The paper closes with a brief discussion of three alternative sets of methodological presuppositions about social reality. Through this analysis, I aim not to renew pointless controversies, for I believe the general linear model (GLM) is a formidable and effective method. But I argue that the model has come to influence our actual construing of social reality, blinding us t important phenomena that can be rediscovered only by diversifying our formal techniques." (Abbot 1988, p. 169)

From a more practical and epistemological perspective, sociologists Pierre Bourdieu and Loic Wacquant, denounced methodological monotheism in both qualitative and quantitative traditions of social research. Indeed, by the time they published their book on reflexivity in social science research, a critical assessment of the hegemony of specific quantitative methods was lacking. Further, they suggest a potential individual-level mechanism for rigid adherence to specific methods that may be related to training traditions and skills acquisition.

> "Rigid adherence to this or that method of data collection will define membership in a "school." the symbolic interactionists being recognizable for instance by the cult of participant observation, ethnomethodologists by their passion for conversation analysis, status attainment researchers by their systematic use of path analysis, etc. And the fact of combining discourse analysis with ethnographic description will be hailed as a breakthrough and a daring challenge to methodological monotheism! We would need to carry out a similar critique in the case of techniques of statistical analysis, be they multiple regression, path analysis, network analysis, or event-history analysis. Here again, with few exceptions, monotheism reigns supreme. Yet the most rudimentary sociology of sociology teaches us that methodological indictments are too often no more than a disguised way of making a virtue out of necessity, of feigning to dismiss, to ignore in an active way, what one is ignorant of in fact" (Bourdieu & Wacquant 1992, p. 226)

In a review article about fertility change, one of the most studied topics in demography and populations studies, Charles Hirschman wrote:

"The standard social science model is that society works pretty much like a regression equation: the task is to find a right set of predictors, solve the equation, and discover what factors are the most important in predicting social outcomes. This framework does lead to empirical generalizations, but there seem to be endless qualifications about the measurement of variables, the meaning and interpretation of variables, the substitutability of one variable for another, and complex interactions with historical settings. If science is to discover parsimonious principles that explain complex patterns, we do not seem to be making progress" (Hirschman 1994, p. 256)

A telling example of the pitfalls of neglecting complex interactions with historical settings has been brought by John Levi Martin and King-To Yeung. The use of the category of race in quantitative American Sociology has led to a "broad but shallow" understanding of racism for three main reasons, two of which relate to the prevalence of regression analysis:

"[...] second, the ease with which controls may be added in regression models; third, the nature of the selective mechanism (gatekeeping such as peer review) in the social sciences, which leads researchers to neutralize as many as possible alternative explanations [...]. This led to an implicit understanding that the goal of sociological research in a racialized society is to "deracialize" its findings" (Martin & Yeung 2003, p. 538)

Tukufy Zuberi, Evelyn Patterson, and Quincy Stewart provided concrete examples of how bivariate and conditional correlations between social outcomes an individual level measures of race are very limited to capture the essential historical and processual aspects of racialized societies, causing a displacement of relevant discussions including the embodied nature of race and the social origin of racial classifications.

"The statistical use of race as an individual, genetic characteristic in debates of race and racial inequality presumes that these latter concepts exist as relationships between variables (e.g., correlations, regression coefficients), not as social processes [...]. This type of analysis imposes a particular social form on race; race is conceived of as either a cause or consequence of other measured characteristics [...] For example, a statistical relationship between IQ and race is presumed to represent either IQ causing race or race causing IQ. The nature of the relationship between these variables may be debated. But the inability of the statistical method—and related research design—to accurately embody race as something more than an individual characteristic is excluded from academic discussion" (Zuberi et al. 2015, p. 119)

More recently, Jennifer Jhonson-Hanks has raised more general concerns regarding the pitfalls of searching for neat effects. Her proposal of a Theory of Conjunctural Action advocates for a new approach to family research that:

> "[...] move us away from seeking to isolate pure effects of specific variables on outcomes and toward understanding how outcomes emerge from the confluence of circumstances. Existing approaches in quantitative social sciences focus on trying to identify exogenous effects, however socially insignificant, at the cost of sometimes ignoring big, real -but endogenous-empirical phenomena" (Johnson-Hanks et al. 2011, p. ix)

*B. List of terms for corpora construction*

We approximate the *ideal risk set* for our study, i.e., the set of papers using statistical methods by selecting two subsets of the *universe*: Corpus 1 and Corpus 2 (see main text for a description). Corpus 1 comprises all articles reporting quantitative methods or data according to the following terms. Research articles reporting methods are classified according to two criteria: First, whether the reported method is a linear model (*linear models*) or other methods (*other methods*). Second, whether they describe general (e.g., linear model) or specific (e.g., Cox regression) methods. In addition, 12 terms refer specifically to frameworks where linear models are used to measure causal relationships, and 15 refer to quantitative data sources.

The lists below include all the 163 terms. Counts printed in parenthesis are the number of unique abstracts in which each term was found. If a term is used several times in the abstract, we only count once. Several terms can be used in the same abstracts. In these cases, we counted them all. Although we lower case all abstracts (and terms we search), we present the original capital letters here to favor the understanding of abbreviations such as Ordinary Least Squares (OLS) and proper names such as Cox regression or model.

1. **Quantitative data** (15 terms): quantitative (973,974), survey (937,519), baseline (632,073), census (43,298), random sample (34,396), panel data (25,489), panel survey (5,345), register data (2,846), vital statistics (2,682), official statistics (1,649), administrative records (1,307), death records (1,187), birth records (1,161), civil registration (755), fiscal data (88).

2. **Linear models - *general*** (48 terms): regression (most strict counting, 1,012,667, flexible counting, 1,056,474), logistic regression (323,728), linear regression (165,232), controlling for (112,083), ANOVA (91,881), analysis of variance (76,157), linear model (59,540), control for (54,474), survival analysis (31,549), Poisson regression (19,586), multivariate model (19,440), generalized linear model (13,712), OLS (12,706), logistic model (11,385), multilevel model (10,227), multinomial logistic regression (9,873), negative binomial (9,779), logit model (8,324), hierarchical model (8,179), ancova (8,021), ordinary least squares (7,369), MANOVA (6,809), hierarchical linear model (4,975), probit model (4,363), time series model (4,212), poisson model (3,419), multinomial regression (2,156), nonparametric regression (2,039), log-linear model (1,867), probit regression (1,492), multi-level model (1,477), mancova (1,193), time-series model (1,108), multilevel regression model (1,024), zero-inflated negative binomial (934), zero-inflated poisson (918), multinomial model (821), gamma model (545), linear probability model (400), multi-level regression model (153), log-log model (140)

3. **Linear models - *specific*** (50 terms): Cox regression (58,989), proportional hazards model (26,350), proportional hazard model (13,610), LASSO (13,202), Cox model (11,397), mixed-effects model (8,452), quantile regression (6,680), random-effects model (6,503), mixed effects model (5,828), random effects model (5,668), generalized additive model (5,138), tobit (3,319), hazard model (3,231), ridge regression (2,644), autoregressive integrated moving average (2,385), mixed-effect model (2,348), hazards model (2,197), mixed effect model (1,975), arima model (1,532), random effect model (1,509), robust regression (1,346), random-effect model (1,346), ordered logit (1,154), age-period-cohort (1,095), competing risk analysis (811), nested models (743), competing risk model (701), competing risks model (415), competing risks analysis (268), ologit (77), SARS model (12), accelerated time failure (5), species-area relationship models (4).

4. **Linear models - *causality*** (12 terms): propensity score matching (15,363), instrumental variable (8,610), difference-in-difference (6,902), randomized control trial (4,177), fixed-effects model (2,126), fixed effects model (1,910), regression discontinuity (1,909), randomized experiment (1,904), propensity-score matching (1,081), fixed-effect model (917), difference in difference (557).

5. **Other methods - *general*** (18 terms): simulation (1,640,662), bayesian (118,087), machine learning (89,695), neural networks (89,395), principal component analysis (76,494), factor analysis (60,575), sequence analysis (57,634), cluster analysis (48,262), deep learning (33,535), network analysis (22,671), correspondence analysis (8,765), social network analysis (5,719), factorial analysis (3,808), multiple correspondence analysis (1,373), social simulation (217), multiple factorial analysis (40), geometric data analysis (21), multivariate descriptive statistics (9).

6. **Other methods - *specific*** (20 terms): discriminant analysis (32,048), Markov chain (31,142), multi-agent (13,394), agent-based (11,557), latent class analysis (5,626), microsimulation (2,559), Gibbs sampling (2,382), qualitative comparative analysis (2,094), Gibbs sampler (1,673), optimal matching (846), individual-based modelling (134), individual-based modeling (114), model-based cluster analysis (113), agent based modeling (57), agent based modelling (44).

Corpus 2 comprises all articles using any combination of words from the following two lists. The first list includes Verbs indicating some type of analysis or interpretation of data, whereas the second one has terms that refer to Evidence or data. A publication is selected into Corpus 2 only when one or more of the words in both lists are present in the abstract.

**Verbs** (5 terms): analysis, analyze, analyse, investigate, study.

**Evidence** (7 terms): data, empirical, evidence, method, methods, model, results.

Combined, these two lists give us an approximation of papers dealing with analyzing or interpreting data regardless of their type (e.g., qualitative or quantitative). Therefore this is a naive approximation of the ideal risk set because it may include qualitative papers. Hence we use this as a conservative set for measuring linear models hegemony.

*C. Finding linear model-based terms in publication's abstracts*

To search abstracts of both corpora for the chosen terms listed above, we first query the Web of Science (WOS) in-house database hosted in PostgreSQL that covers publications from 1990 up to the end of April of 2022. We limit the publications' document type to "articles", then lowercase the abstract and trim the potential starting and ending whitespaces. Next, we use PostgreSQL "like" function with wildcards for *exact* text matching while allowing the text to be more extended *before* or *after* the intended term combination, but the exact word combination needs to have been used in the text (for instance, we search for "LOWER(TRIM(abstract.TEXT)) like '%cox model%'" which will also include "cox models" in the results. It will not allow for a change in the first term of the bigram of terms. For example, "cox's model" will not be included in the sample).

Using this strategy, for Corpus 1, we search for every *general* or *specific* term in linear model-based research or other methods (listed above) in the abstract. For Corpus 2, we search for a combination of the terms in the two arrays (Verbs and Evidence) to co-occur once or multiple times in the same abstract.

In the next step and for both selected Corpora, using Python 3 base (https://docs.python.org/3/library/) and re (https://docs.python.org/3/library/re.html) libraries, we use the lowercase abstract text and find all occurrences of these terms using a regex pattern defined and compiled as "r'(?:%s)'" (replacing "%s" with our terms) that would return any usages of the terms or term combinations in the abstract text (while respecting the sentence structure, for instance, a sentence ending in one word and the next sentence starting with the second word in the bigram will be excluded, e.g., "This model is called cox. Model needs to be initiated." will not be selected as an instance of using "cox model" since there is a dot "." used between "cox" and "model"). As a further robustness control, and for the case of "regression" and "OLS", in a parallel attempt, we search using "r'\b(?:%s)\b'" regex pattern that considers the word boundaries and returns the exact once or more usages of "regression" or "OLS" in the abstract (example Python 3 code below).

In addition, we should emphasize that our chosen 163 term combinations overlap with each other to cover all possible use-cases (e.g., "autoregressive integrated moving average" and "autoregressive integrated moving average model" have only one-word difference, "model" used at the end) and our search favors the *least extended* term combination (i.e., a shorter subset of term combination) and in the usage counts, we present these as separate counts hence some term combinations do not show up. This strategy enables us to cover **different naming and labeling used by subdisciplines or communities**. It is problematic to count one form more than or instead of the other. Still, since we are focused on the "group" of terms as treatment and control (e.g., linear model specific versus other methods), this does not affect our results because we are not comparing terms in one category with each other. For specific cases of "regression" and "OLS", we exclude them from more flexible search function results and instead present the count of most strict use-cases while considering word boundaries (for "regression", we also present the *flexible* use counts for comparison purposes).

==== Example Python code ====

```python
import re

def search_only_regression(x):
    keywords = [r'regression']

    p = re.compile(r'\b(?:%s)\b' % '|'.join(keywords), flags=re.IGNORECASE)
    try:
        res2return = p.findall(x)
        if any(res2return):
            return [x.lower() for x in res2return if x]
        else:
            return None
    except TypeError:
        return None

def search_only_regression_flex(x):
    keywords = [r'regression']

    p = re.compile(r'(?:%s)' % '|'.join(keywords), flags=re.IGNORECASE)
    try:
        res2return = p.findall(x)
        if any(res2return):
            return [x.lower() for x in res2return if x]
        else:
            return None
    except TypeError:
        return None
```

*D. Complementary results and figures*

Figure A1 displays the temporal trend in the proportion of articles reporting linear models and other methods according to the type of word used (i.e., *general*, *specific*, or *causality*-related) for the six OECD macro fields of science.

**Fig. A1.**

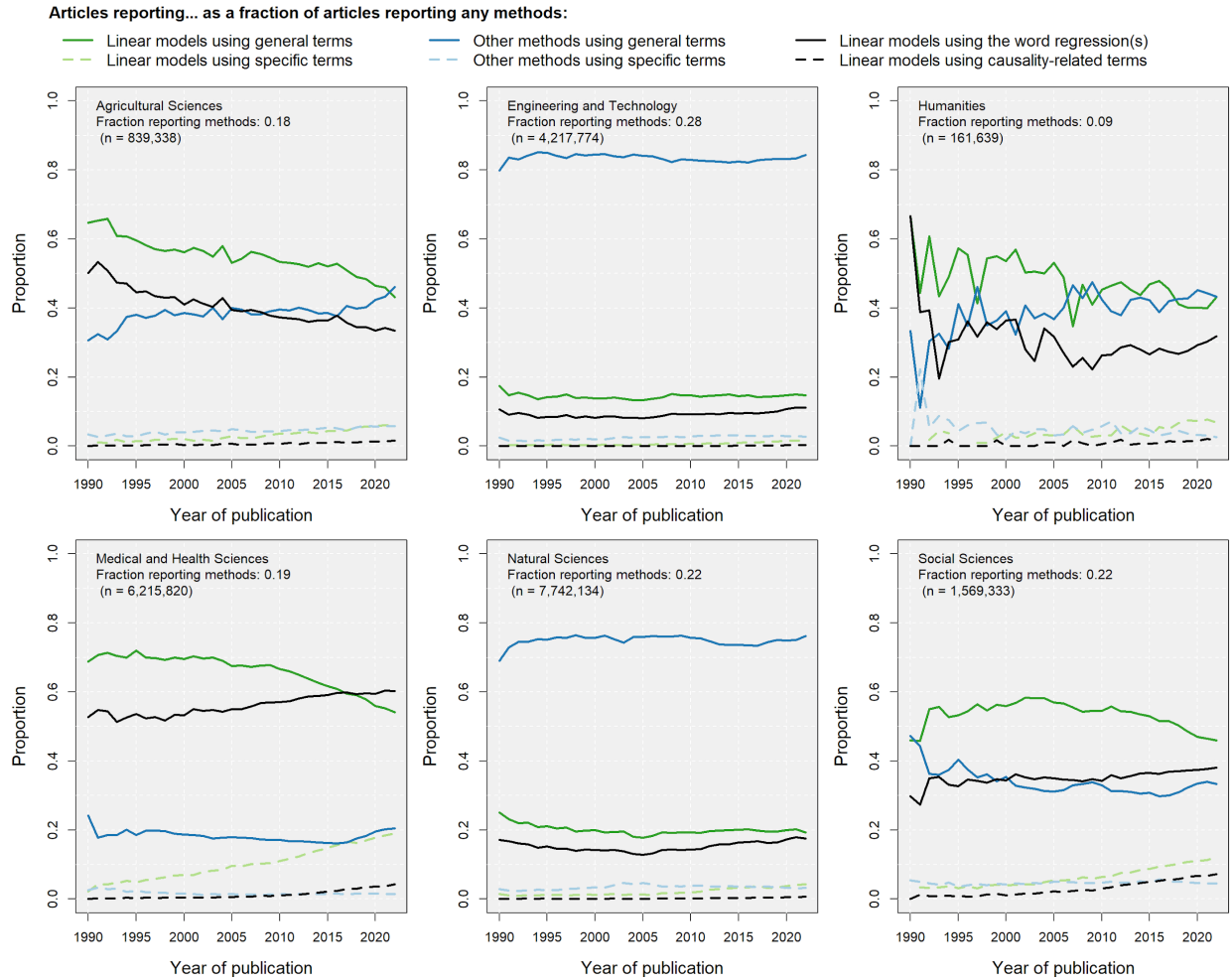

**The most prevalent type of method (linear models vs. *other methods*) is typically reported using general terms.** Temporal trend in the distribution of articles reporting statistical methods according to the type of method reported (linear model or other methods) and the type of term (general, specific, causality-related).

**Fig. A2.**

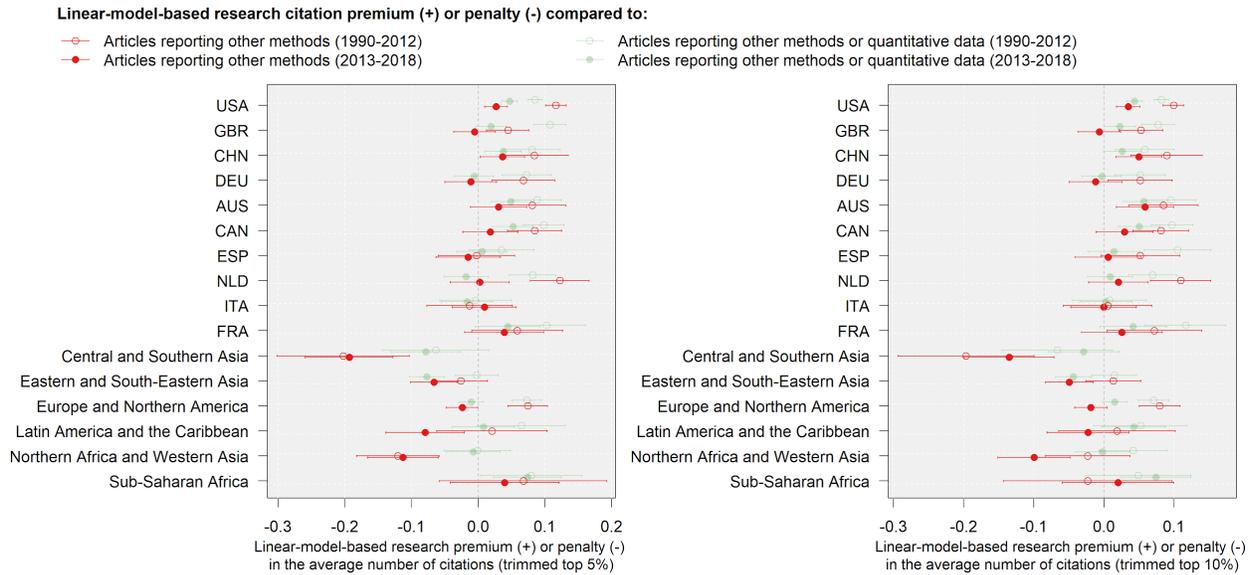

**The visibility advantage of *linear-model-based research* compared to quantitative research using other methods and quantitative data sources.** Log-scaled differences in the odds of having at least one citation (left panel) and the number of citations (right panel) in the first three years after publication between (i) articles reporting linear models and articles reporting other methods (red dots and lines) and (ii) articles reporting other methods or quantitative data (light green dots and lines). Regressions are stratified by countries with the ten most considerable contributions to articles in the Social Sciences (USA: the United States of America, CHN: China, GBR: the United Kingdom, DEU: Germany, NDL: the Netherlands, FRA: France, CAN: Canada, ITA: Italy, ESP: Spain, and AUS: Australia) and the UN-SDG regions for the other countries and by the median year of publication (i.e., 2011). Australia and New Zealand are included in Europe and Northern America, and other countries from Oceania are excluded due to the small sample size (1,350 articles).

**Table A1.**
List of all used terms as a table with groupings

| Number | Term combination | Type of term | Group of term |
|---|---|---|---|
| 1 | accelerated time failure | specific | treatment |
| 2 | accelerated time failure model | specific | treatment |
| 3 | administrative records | data type | quantitative |
| 4 | agent based modeling | specific | control |
| 5 | agent based modelling | specific | control |
| 6 | agent-based | specific | control |
| 7 | agent-based modeling | specific | control |
| 8 | agent-based modelling | specific | control |
| 9 | age-period-cohort | specific | treatment |
| 10 | age-period-cohort model | specific | treatment |
| 11 | analyse | action | array 2 |
| 12 | analysis | action | array 2 |
| 13 | analysis of variance | general | treatment |
| 14 | analyze | action | array 2 |
| 15 | ancova | general | treatment |
| 16 | anova | general | treatment |
| 17 | arima model | specific | treatment |
| 18 | autoregressive integrated moving average | specific | treatment |
| 19 | autoregressive integrated moving average model | specific | treatment |
| 20 | baseline | data type | quantitative |
| 21 | bayesian | general | control |

| Number | Term combination | Type of term | Group of term |
|---|---|---|---|
| 22 | birth records | data type | quantitative |
| 23 | census | data type | quantitative |
| 24 | civil registration | data type | quantitative |
| 25 | cluster analysis | general | control |
| 26 | competing risk analysis | specific | treatment |
| 27 | competing risk model | specific | treatment |
| 28 | competing risks analysis | specific | treatment |
| 29 | competing risks model | specific | treatment |
| 30 | control for | general | treatment |
| 31 | controlling for | general | treatment |
| 32 | correspondence analysis | general | control |
| 33 | cox model | specific | treatment |
| 34 | cox regression | specific | treatment |
| 35 | cox regression model | specific | treatment |
| 36 | data | piece of evidence | array 1 |
| 37 | death records | data type | quantitative |
| 38 | deep learning | general | control |
| 39 | difference in difference | causality | treatment |
| 40 | difference-in-difference | causality | treatment |
| 41 | discriminant analysis | specific | control |
| 42 | empirical | piece of evidence | array 1 |
| 43 | evidence | piece of evidence | array 1 |
| 44 | factor analysis | general | control |
| 45 | factorial analysis | general | control |

| Number | Term combination | Type of term | Group of term |
|---|---|---|---|
| 46 | fiscal data | data type | quantitative |
| 47 | fixed effects model | causality | treatment |
| 48 | fixed-effect model | causality | treatment |
| 49 | fixed-effects model | causality | treatment |
| 50 | gamma model | general | treatment |
| 51 | generalized additive model | specific | treatment |
| 52 | generalized linear model | general | treatment |
| 53 | geometric data analysis | general | control |
| 54 | gibbs sampler | specific | control |
| 55 | gibbs sampling | specific | control |
| 56 | hazard model | specific | treatment |
| 57 | hazards model | specific | treatment |
| 58 | hierarchical linear model | general | treatment |
| 59 | hierarchical model | general | treatment |
| 60 | individual-based modeling | specific | control |
| 61 | individual-based modelling | specific | control |
| 62 | instrumental variable | causality | treatment |
| 63 | investigate | action | array 2 |
| 64 | lasso | specific | treatment |
| 65 | latent class analysis | specific | control |
| 66 | linear model | general | treatment |
| 67 | linear probability model | general | treatment |
| 68 | linear regression | general | treatment |
| 69 | logistic model | general | treatment |

| Number | Term combination | Type of term | Group of term |
|---|---|---|---|
| 70 | logistic regression | general | treatment |
| 71 | logit model | general | treatment |
| 72 | log-linear model | general | treatment |
| 73 | log-log model | general | treatment |
| 74 | machine learning | general | control |
| 75 | mancova | general | treatment |
| 76 | manova | general | treatment |
| 77 | markov chain | specific | control |
| 78 | markov chain monte carlo | specific | control |
| 79 | method | piece of evidence | array 1 |
| 80 | methods | piece of evidence | array 1 |
| 81 | microsimulation | specific | control |
| 82 | mixed effect model | specific | treatment |
| 83 | mixed effects model | specific | treatment |
| 84 | mixed-effect model | specific | treatment |
| 85 | mixed-effects model | specific | treatment |
| 86 | model | piece of evidence | array 1 |
| 87 | model-based cluster analysis | specific | control |
| 88 | multi-agent | specific | control |
| 89 | multi-agent-based modeling | specific | control |
| 90 | multi-agent-based modelling | specific | control |
| 91 | multilevel model | general | treatment |
| 92 | multi-level model | general | treatment |
| 93 | multilevel regression model | general | treatment |

| Number | Term combination | Type of term | Group of term |
|---|---|---|---|
| 94 | multi-level regression model | general | treatment |
| 95 | multinomial logistic regression | general | treatment |
| 96 | multinomial model | general | treatment |
| 97 | multinomial regression | general | treatment |
| 98 | multinomial regression model | general | treatment |
| 99 | multiple correspondence analysis | general | control |
| 100 | multiple factorial analysis | general | control |
| 101 | multivariate descriptive statistics | general | control |
| 102 | multivariate model | general | treatment |
| 103 | negative binomial | general | treatment |
| 104 | negative binomial model | general | treatment |
| 105 | nested models | specific | treatment |
| 106 | network analysis | general | control |
| 107 | neural networks | general | control |
| 108 | nonparametric regression | general | treatment |
| 109 | official statistics | data type | quantitative |
| 110 | ologit | specific | treatment |
| 111 | ols | general | treatment |
| 112 | optimal matching | specific | control |
| 113 | ordered logit | specific | treatment |
| 114 | ordinary least squares | general | treatment |
| 115 | ordinary least squares regression | general | treatment |
| 116 | panel data | data type | quantitative |
| 117 | panel survey | data type | quantitative |

| Number | Term combination | Type of term | Group of term |
|---|---|---|---|
| 118 | poisson model | general | treatment |
| 119 | poisson regression | general | treatment |
| 120 | principal component analysis | general | control |
| 121 | probit model | general | treatment |
| 122 | probit regression | general | treatment |
| 123 | propensity score matching | causality | treatment |
| 124 | propensity-score matching | causality | treatment |
| 125 | proportional hazard model | specific | treatment |
| 126 | proportional hazards model | specific | treatment |
| 127 | qualitative comparative analysis | specific | control |
| 128 | quantile regression | specific | treatment |
| 129 | quantitative | data type | quantitative |
| 130 | random effect model | specific | treatment |
| 131 | random effects model | specific | treatment |
| 132 | random sample | data type | quantitative |
| 133 | random-effect model | specific | treatment |
| 134 | random-effects model | specific | treatment |
| 135 | randomized control trial | causality | treatment |
| 136 | randomized experiment | causality | treatment |
| 137 | randomized-control-trial | causality | treatment |
| 138 | register data | data type | quantitative |
| 139 | regression | general | treatment |
| 140 | regression analysis | general | treatment |
| 141 | regression discontinuity | causality | treatment |

| Number | Term combination | Type of term | Group of term |
|---|---|---|---|
| 142 | regression model | general | treatment |
| 143 | results | piece of evidence | array 1 |
| 144 | ridge regression | specific | treatment |
| 145 | robust regression | specific | treatment |
| 146 | sars model | specific | treatment |
| 147 | sequence analysis | general | control |
| 148 | simulation | general | control |
| 149 | social network analysis | general | control |
| 150 | social simulation | general | control |
| 151 | species area relationship models | specific | treatment |
| 152 | species-area relationship models | specific | treatment |
| 153 | study | action | array 2 |
| 154 | survey | data type | quantitative |
| 155 | survival analysis | general | treatment |
| 156 | time series model | general | treatment |
| 157 | time-series model | general | treatment |
| 158 | tobit | specific | treatment |
| 159 | vital statistics | data type | quantitative |
| 160 | zero-inflated negative binomial | general | treatment |
| 161 | zero-inflated negative binomial model | general | treatment |
| 162 | zero-inflated poisson | general | treatment |
| 163 | zero-inflated poisson model | general | treatment |